\documentclass[prl,twocolumn,showpacs,showkeys,preprintnumbers,amsmath,amssymb]{revtex4}

\usepackage{graphicx}
\usepackage{dcolumn}
\usepackage{bm}

\def\duzomniejsze{<\kern-.7mm<}
\def\duzowieksze{>\kern-.7mm>}

\def\textbf#1{{\bf #1}}
\def\beq{\begin{equation}}
\def\eeq{\end{equation}}
\def\be{\begin{equation}}
\def\ee{\end{equation}}
\def\ben{\begin{eqnarray}}
\def\een{\end{eqnarray}}
\def\beqa{\begin{eqnarray}}
\def\eeqa{\end{eqnarray}}
\def\eea{\end{array}}
\def\bea{\begin{array}}
\newcommand{\bei}{\begin{itemize}}
\newcommand{\eei}{\end{itemize}}
\newcommand{\bee}{\begin{enumerate}}
\newcommand{\eee}{\end{enumerate}}

\newcommand{\p}{p} 

\usepackage{xcolor}
\newcounter{mnotecount}[section]
\renewcommand{\themnotecount}{\thesection.\arabic{mnotecount}}
\newcommand{\mnote}[1]
{\protect{\stepcounter{mnotecount}}$^{\mbox{\footnotesize
$
\bullet$\themnotecount}}$ \marginpar{
\raggedright\tiny\em
$\!\!\!\!\!\!\,\bullet$\themnotecount: #1} }

\usepackage{soul}

\begin{document}


\title{The double doors of the horizon}

\author{Erik Aurell}
\email{eaurell@kth.se}
\affiliation{KTH -- Royal Institute of Technology, AlbaNova University Center, SE-106 91 Stockholm, Sweden}%

\date{\today}

\begin{abstract}
In statistical mechanics
entropy is a measure of 
disorder obeying Boltzmann's formula $S=\log{\cal N}$, where ${\cal N}$ is the accessible phase space volume.
In black hole thermodynamics one associates to a black hole an entropy Bekenstein-Hawking $S_{BH}$.
It is well known that $S_{BH}$ is very large for astrophysical black holes, much larger than
any collection of material objects that could have given rise to the black hole.
If $S_{BH}$ is an entropy the question is thus what is the corresponding ${\cal N}$, and how come this
very large phase space volume is only opened up to the universe by a gravitational collapse,
which from another
perspective looks like a massive loss of possibilities. 
I advance a hypothesis that the very large increase in entropy can perhaps
be understood 
as an effect of classical gravity, which eventually bottoms out when quantum gravity comes into play.
I compare and discuss a selection of the very rich literature around these questions.
\end{abstract}

\pacs{03.67.Lx, 42.50.Dv}
\maketitle

{\it Introduction:} The Bekenstein-Hawking
entropy $S_{BH}$ of a black hole is a  
central
quantity in modern physics.
In its general form $S_{BH}$ equals $\frac{1}{4}\frac{A}{l_{\p}^2}$
where $A$ is the surface area of the black hole horizon, and 
$l_{\p}\approx 1,6\cdot 10^{-35}\,  \hbox{m}$
is the Planck length.
This form of the entropy formula includes extreme cases 
where the black hole is rotating 
at high speed and/or is very strongly electrically charged, and where 
$A$ can be arbitrarily small.
In the simplest setting of a
non-rotating electrically neutral black hole 
(Schwarzschild geometry) the area of the horizon is 
$16\,\pi\, l_{\p}^2\, \frac{M^2}{m_{\p}^2}$,
where $M$ is the mass of the object and
$m_{\p}\approx 2\cdot 10^{-8}\, \hbox{kg}$  is the Planck mass.  
In this case, roughly appropriate for all black holes that have been
observed to date, $S_{BH}$ equals $4\pi\, \frac{M^2}{m_{\p}^2}$
and increases faster with mass 
than any aggregate of particles and fields that could have 
formed the black hole~\cite{Bekenstein1973,tHooft93}.
Consequently, for large enough mass, $S_{BH}$
will be larger than than the entropy of any astrophysical object 
that could have formed the black hole.
Well-known estimates say that the entropy of star
like the sun would increase about $10^{19}$ times if it
could collapse into a black hole.
In the total entropy budget of the universe 
almost all the entropy is held by super-massive black holes~\cite{EganLineweaver2010}.

From the point of view of statistical mechanics
entropy is a measure of 
disorder obeying Boltzmann's formula $S=\log{\cal N}$.
Disorder means that the system can only be described up
to a phase space volume ${\cal N}$.
While the formation of an event horizon
closes the door 
on some types of classical evolution, 
since a classical observer can never get out of a black hole,
in some other sense a gravitational collapse must open
a door to new possibilities;
"Opened are the double doors of the horizon // unlocked are its bolts" \cite{Glass}.

In this note I put forward 
a hypothesis that
the increase of disorder
is due to the classical chaotic dynamics of a sufficiently strong gravitational field. 
The existence of 
such a mechanism is plausible, 
and supported by well-established theories, to be
referenced below.
The mechanism has nevertheless 
not previously been
put forward as a physical mechanism which can generate $S_{BH}$.

{\it The relation to Hawking's black hole information paradox:}
Bekenstein-Hawking entropy 
is at the center of a very extensive literature 
motivated by Hawking's black hole information paradox~\cite{Hawking76,Preskill92,Verlinde-Verlinde2013,Harlow16,UnruhWald17}.
Possible solutions to the information paradox
were in \cite{Giddings}
 classified as (i) fundamental information loss, (ii) remnants or new physics at the horizon, and (iii) information return in Hawking radiation.
I am here only concerned with the last scenario, first proposed by Page~\cite{Page80}, and more recently 
dubbed the "central dogma" of black hole quantum physics~\cite{Shaghoulian21}.
Quantum mechanics is thus assumed to hold everywhere
in the universe, including in the interior of black holes.
Some (unknown) unitary operator
evolves an initially pure state of matter to a gravitational collapse
and then onto Hawking radiation escaping to infinity.
In \cite{FOP2021,II} we considered the kinematics of such a process, where
in the final state radiation will be in multi-mode entangled state.
That such multi-party entanglement should be present in 
Hawking radiation is known already from \cite{Page80},
where it was pointed out that when one Hawking particle is 
emitted with 
some momentum, the remaining black hole must carry the opposite
momentum. Hence later Hawking particles must be entangled
with earlier Hawking particles, though only weakly by 
this recoil effect.

A distinction now has to be made between entropy in the informational-theoretic sense, and entropy in the thermodynamic sense. Entropy of the first kind, usually called "fine-grained entropy" in the black hole literature, does not change under unitary evolution. Entropy of the second kind, usually called "coarse-grained entropy" in the black hole literature,
will however tend to
a maximum at given external constraints. A part of a resolution 
of the information paradox of type  
(iii) must hence be that 
information-theoretic entropy is
preserved in unitary evolution, 
and that $S_{BH}$ is a thermodynamic
entropy, an observation also due to 
Hawking \cite{Hawking14}.
Another argument that 
$S_{BH}$ is a thermodynamic entropy
follows from comparing 
$S_{BH}$  to
the entropy 
of black-body radiation emitted by an evaporating black hole, mode by mode.
This can be estimated from the law of Stefan-Boltzmann and its 
generalizations  
\cite{Zurek1982,Page1983,Page2005}, and is somewhat larger than $S_{BH}$, as appropriate 
for an irreversible process. The ratio is however a constant
independent of mass, meaning
that on a relative scale almost all the entropy increase happens from when
matter starts collapsing until when the resulting black hole has settled 
down to a stationary state. 

In the quantum domain the role of information-theoretic entropy
is taken by von Neumann entropy, which conserved for a closed
system, but can change for a subsystem.
Important progress was in the last less than ten years made in describing a black hole as a unitary quantum system with $\exp\left(\frac{A}{4 l_{\p}^2}\right)$ states
\cite{Lewkowycz13,Engelhardt15,Almheiri19,Penington19,Almheiri20}.
This approach allowed for the first time a calculation of the 
von Neumann entropy
of Hawking radiation, showing that it
first increases, as does the thermodynamic entropy (as above),
but then decreases, following the Page curve.
This approach however does not deal with the degrees of freedom of the space-time geometry
in its entirety, including the full interior of the black hole~\cite{Maldacena21}.
The question of how thermodynamic entropy can increase so
dramatically in a collapse to a black hole,
a process which can be taken to happen before any Hawking particles
have been emitted at all, has also not been addressed 
in this important series of papers. 

{\it Black holes quantum cores, chaos and space-time chaos:}
I now introduce three assumptions of a physical 
or plausibility nature,
patterned after the "naive model" 
of \cite{rovelli2014planck}, see also \cite{paik2018black}:
(A) after the gravitational collapse has run its course there emerges in the
center of the black hole a quantum core  
state of matter and gravitation;
(B) the spatial support of this state 
is about $r_c\sim l_{\p}\,\left(\frac{M}{m_{\p}}\right)^{\frac{1}{3}}$
such that the density of the core is about Planck density;
(C) outside the core space-time is a meaningful concept, and to a good approximation given by a metric which obeys Einstein's equations.
All the mass that went down into the black hole
is assumed concentrated in a volume of size $r_c$, and thoroughly transformed
by and entangled with the gravitational field.
For a solar mass black hole $r_c\sim  10^{13}\, l_{\p}\approx 10^{-22}\,\hbox{m}$.

On a general level, during the collapse process
the space-time inside a
physical black hole can be assumed to be
both unique and complex, where in classical theory 
increasing gravity would 
eventually lead to a singularity in a way that reflects the black hole’s formation~\cite{Uggla2013}.
One can try to estimate using
classical concepts the increase of disorder when a body falls down in the black hole before it hits the core. 
First, it is reasonable to take any such body an elementary particle,
as macroscopic bodies are torn apart by tidal forces
well before hitting the final singularity
\footnote{
On a body of size $l$ and mass $m$ at a distance $r$ from a black hole of mass $M$ ($l \ll r$ the tidal forces are about $GmMl/r^3$ 
Imagine a
system of size $l$ and binding energy $V$
held together by a force about $V/l$;
it is then torn apart at a distance
from the singularity of about
$r_*\sim r_s \left(\frac{l^2}{r_s^2} \frac{m c^2}{V} \right)^{\frac{1}{3}}$.
For a hydrogen atom in a solar mass black hole 
$r_*$ is about one millimeter 
($10^{-6}r_s$), as follows from
$l\sim 10^{-10}$m, $r_s\sim 10^{3}$m, 
$m c^2\sim 10^9$eV and $V\sim 13 $eV.
A proton would similarly be torn apart at a distance of
one nanometer from the singularity ($10^{-12}r_s$), as follows from $l\sim 10^{-15}$m and $V\sim m c^2$, though one may imagine that pulling apart the quarks could also lead 
to particle production and a transition to another phase.
In any case, both distances are small compared to $r_s$ but very large compared to $r_c$.}.
Second, the same diverging tidal forces mean that a small difference between
a reference geodesics and a deviation grows without limit as the reference
geodesics approaches the singularity. This 
type of chaos leads to a growth which is algebraic
in the distance from the singularity at the final time, \textit{see}
\cite{MisnerThorneWheeler} (chap. 32.6) or \cite{bazanski1989geodesic},
which in turn translates to a power-law of the ratio $\frac{M}{m_{\p}}$.
The number of bits needed to specify an initial condition leading to
a given final condition is thus logarithmic in $\frac{M}{m_{\p}}$.
Chaos of the geodetic motion of the body falling
down the hole can therefore only correspond to a 
sub-leading fraction of the entropy increase.

Third, disorder can also emerge
from the
chaotic non-linear dynamics of the gravitational field itself.
This mechanism (BKL scenario) was first proposed 
for homogeneous cosmology~\cite{BKL,Khalatnikov1985,BKL1982},
and has more recently been conjectured
to also describe aspects of the
final stages of the collapse of an astrophysical 
body~\cite{Heinzle2012}.
For chaotic dynamical systems (coherent in space)
the increase in entropy per unit time is measured 
the Kolmogorov-Sinai entropy.
For the homogeneous BKL solution KS-entropy
(KS-entropy of the Gauss map)
is known, and is positive~\cite{Khalatnikov1985}.
This should however again only correspond to a 
sub-leading fraction
of the entropy increase.
However, in BKL, every matter particle 
loses causal contact with every other matter particle, a property known as asymptotic silence~\cite{Andersson2005}.
Therefore, one would not expect the gravitational field around
different points to be synchronized, and 
the general BKL-like solution has consequently
been postulated to be
inhomogeneous (turbulent) with different realizations
of the homogeneous solution around different spatial points \cite{Belinskii1992}.
If this mechanism of space-time chaos
can produce the increase of phase space volume quantified 
Bekenstein-Hawking entropy is hence the question.

{\it Stability and instability of Schwarzschild and geometry:}
The exterior of the The Schwarzschild black hole was shown to be linearly stable 
on the physical level of rigor  
in the 1950ies and 1960ies
\cite{ReggeWheeler1957,Vishveshwara1968}; mathematical proofs are quite recent
\cite{Dafermos2019,Johnson2019}.
Very recently a mathematical proof of nonlinear stability was also presented 
in~\cite{Dafermos2021}, unfortunately not very accessible to physicists
\footnote{This so far unpublished monograph is 519 pages in length}.
In another line of investigations from the same community, dynamics in the 
interior region has also been considered
from the rigorous point of view \cite{dafermos2005interior,DafermosLuk2017}
Two further papers have been announced, but have so far not appeared.

This (rigorously established) stability 
does not exclude that in the interior there can be the kind of instabilities which in plasma physics and hydrodynamics
are called convective~\cite{Bers1975,HuerreMonkewitz1985}. Imagine a perturbation of the metric at some distance $r$ from the smeared singularity. Such a perturbation can decay at the point $r$ where it is introduced (absolute stability), but at the same time grow, for a finite or infinite time, at smaller $r$. In a classical model this growth can be unlimited by moving to ever smaller $r$, but by the assumptions made here it would stop at $r_c$. Unfortunately, the disorder created by such an instability seems quite hard to estimate.

In a quantum theory of gravity the information would instead be stored in the state at the smeared singularity. The ensemble describing all such space-times, assuming they will in the end be thoroughly mixed and indistinguishable from the outside, must be able to encode all quantum states of all systems that could have given rise to the black hole of given mass, angular momentum and electric charge.
By one of Bekenstein's original estimates such
an ensemble has an entropy of the order of $S_{BH}$
\cite{Bekenstein1973}.
More refined estimates in the same direction were more
recently given by Mukhanov \cite{Mukhanov2003}.
Hence, one can at least say that the classical chaos of the gravitational field
should not
lead to an entropy increase \textit{larger} than $S_{BH}$.

\section*{Acknowledgments}
The idea that the increase of entropy in a graviational collapse is related to classical chaos was raised by Prof Bo Sundborg in discussions with the author.
I thank Angelo Vulpiani for many enlightening discussions,
Pawe{\l} Horodecki and Micha{\l} Eckstein for a previous collaboration
which led to this work,
and Claes Uggla and Ingemar Bengtsson for 
helpful and constructive remarks.
This work was supported by Swedish Research Council grant 2020-04980.

\bibliography{QBH}

\section*{Remarks on classical and quantum entropy increase and thermalization:}
The mechanism proposed above is based on the distinction
between information-theoretical entropy and thermodynamic 
entropy.
Thermodynamic entropy can 
in turn be interpreted as
lack of information (ignorance, or coarse-graining)~\cite{Szilard25,Mandelbrot62,Jaynes57a,Jaynes57b},
or as a property
of an ensemble describing the outcome of
possible (sufficiently simple) experiments after a process of thermalization \cite{Khinchin,Lebowitz1999}.

If and how a closed quantum system thermalizes (or does not)
is an important topic, often referred to as the 
Eigenvalue Thermalization Hypothesis (ETH)
\cite{Srednicki1994,DAllessio2016}.
In quantum chaos growth of disorder can often be 
described by classical dynamics which lead to 
generation of smaller and smaller structures in phase space, until quantum effects
eventually take over \cite{Berry1979}.
The classical picture proposed here is therefore not
necessarily in contradiction with the idea that 
the quantum state in the center of a black hole 
is quantum chaotic and in itself scrambles information 
in a fast way \cite{Sekino2008}.
A precise mechanism in this direction, however assuming the holographic
principle, was discussed in \cite{lowe2022quantum}.

\section*{Remarks on Bekenstein's 2001 critique of his 1973 interpretation of BH entropy} 
Above I evoked Bekenstein's argument in his pioneering 1973 paper that
$S_{BH}$ is the
number of quantum states of matter than could
have formed a black hole~\cite{Bekenstein1973}.
I thus have to address the 
counter-argument advanced by Bekenstein in a 2001 
review~\cite{Bekenstein2001}.
Bekenstein there
considered a black hole emitting Hawking radiation
and at the same time being fed by a stream of matter 
so that its mass, angular momentum and charge stay constant.
Bekenstein formulated his objection 
(page 8 in~\cite{Bekenstein2001}) as follows:
\begin{quote}
\textit{[The black hole then] does not change in time, and neither does its entropy. But surely the inflowing matter is bringing into the black hole fresh quantum states; yet this is not reflected in a growth of $S_{BH}$! 
[...] If we continue thinking of the Hawking radiation as originating outside the horizon, this does not sound possible.}
\end{quote}
A part of a counter-argument runs as what Bekenstein
outlines is not a state of thermal equilibrium but a
non-equilibrium stationary state (NESS). 
An entropy flow can be defined as the expected loss of entropy
of the stream of infalling matter, or as the expected gain of entropy
of the outgoing Hawking radiation: in a stationary state these
should match. A mechanism supporting such an entropy flow was discussed
by the author and Micha{\l} Eckstein and Pawe{\l} Horodecki in \cite{FOP2021}.

Another part of the counter-argument is that the entropy brought into the black hole
by fresh quantum matter is anyway only an extremely small fraction of all the entropy
in play in the process. The increase of
Bekenstein-Hawking entropy of the black hole when its mass increases by $\Delta m$ is
$8\pi \frac{M}{m_{\p}^2}\delta m$. This is the same as if an impacting body of 
mass $\Delta m$ would carry an entropy per unit mass of $c^2 /k_B T_H$ where $T_H$
is the Hawking temperature of the black hole. For a solar mass black hole this is 
about $10^{47}\cdot \hbox{kg}^{-1}$.
By comparison, the standard molar entropy of water at room temperature
is $75.33 \hbox{J/(mol K)}$ which means about $3\cdot 10^{26}\cdot \hbox{kg}^{-1}$
in the same units.
To emphasize this discrepancy in entropy per mass of more than twenty orders of magnitude
was the main motivation of this note.

\end{document}